\documentstyle[12pt]{article}

\def\be{\begin{equation}}
\def\ee{\end{equation}}
\def\a{\alpha}
\def\b{\beta}
\def\l{\lambda}

\def\lim{\mbox{lim}}
\def\ra{\rangle}
\def\la{\langle}
\def\tr{\mbox{Tr}}

\begin{document}

\begin{center}
{\bf Baxter Q- operator and functional relations.} 
\end{center}
\vspace{0.2in}
\begin{center}
{\large A.A.Ovchinnikov}
\end{center}

\begin{center}
{\it Institute for Nuclear Research, RAS, Moscow}
\end{center}

\vspace{0.1in}

\begin{abstract}

We obtain the Baxter Q-operators in the $U_q(\widehat{sl}_2)$ 
invariant integrable models as a special limits of 
the quantum transfer matrices corresponding to different 
spins in the auxiliary space both from the functional relations 
and from the explicit calculations.  
We derive the Baxter equation from the well known fusion 
relations for the transfer matrices. 
Our method is valid for an arbitrary integrable model 
corresponding to the quantum group $U_q(\widehat{sl}_2)$
for example for the XXZ- spin chain.

\end{abstract}

\vspace{0.2in}

{\bf 1. Introduction}

\vspace{0.1in}

At present time the quantum groups and the universal $R$- matrix 
\cite{Jimbo}, \cite{Drinfeld} are the powerful tools to study 
various aspects of quantum integrable models such as the 
Yang-Baxter equation, Baxter $Q$- operator and functional relations. 
Using the special q-oscillator representations of the quantum 
group $U_q(\widehat{sl}_2)$ for the auxiliary space the Baxter 
$Q$- operator \cite{Baxter} was constructed for the Conformal 
Field Theory \cite{BLZ} and the six-vertex model \cite{F}, \cite{J}. 
Using the explicit form of the universal $R$- matrix \cite{KT} 
the $L$- operator for the Baxter $Q$- operator was derived in 
\cite{N1}. The functional relations in the universal form for 
the $U_q(\widehat{sl}_2)$- invariant integrable models were 
obtained in \cite{N2}. 
In the present paper we continue to study the Baxter $Q$- operator 
in various $U_q(\widehat{sl}_2)$- invariant integrable models. 
We discuss the connection between the $j\rightarrow\pm\infty$ 
limits of the quantum transfer matrix $Z_j$ and the Baxter $Q$- 
operators. First, we derive the correspondence between the 
limits of the spin-$j$ representations of the quantum group and 
the q-oscillator representations. Second, we find the limits 
of the transfer matrices and show that it is given by the 
$Q$- operators. 
We derive these results both using the functional relations 
and by the explicit microscopic calculations. 
The last method seems to be an interesting and new contribution 
to the subject. In fact, only the microscopic calculations 
can be the basis for the complete understanding of the whole 
set of the limits of various transfer matrices. 
The calculations are far from being trivial and have not been 
presented in the literature. 
Finally, we obtain the Baxter equation 
as a limit of the functional relations in the universal form 
(fusion relations). Thus we show that the $Q$- operator could be 
derived without the knowledge of the q-oscillator representations.

\vspace{0.2in}

{\bf 2. Quantum group $U_q(\widehat{sl}_2)$. }

\vspace{0.1in}

We begin with the definition of the algebra $U_q(\widehat{sl}_2)$ 
and the corresponding universal $R$- matrix. 
It is the complex associative algebra with the generators 
corresponding to the simple roots $e_i,f_i,h_i$, $i=0,1$ 
obeying the following defining relations: 
\[
[h_i,h_j]=0,~~~[h_i,e_j]=a_{ij}e_j,~~~ [h_i,f_j]=-a_{ij}f_j, ~~~ 
[e_i,f_j]=\delta_{ij}\frac{q^{h_i}-q^{-h_i}}{q-q^{-1}}, 
\]
where $a_{ij}$ is the Cartan matrix $A$ which for the case 
of the algebra $\widehat{sl}_2$ equals 
\[
A=\left(
\begin{array}{cc}
2 & -2 \\
-2 & 2 \\
\end{array}\right). 
\]
This definition should be supplemented by the Serre relations 
satisfied for all distinct $i$ and $j$: 
\[
e_i^{3}e_j-[3]_q e_i^{2}e_j e_i+[3]_q e_i e_j e_i^{2}-e_j e_i^{3}=0,  
\]
\[
f_i^{3}f_j-[3]_q f_i^{2}f_j f_i+[3]_q f_i f_j f_i^{2}-f_j f_i^{3}=0. 
\]  
We did not need the grading element $d$ in the present paper. 
The generators $e_i,f_i,h_i$ correspond the simple roots 
vectors $\a$, $\b$ according to the rule 
$e_0=e_{\b}$, $e_1=e_{\a}$ and the same for $f_i,h_i$. 
Define the root vector $\delta=\a+\b$. Then the root lattice 
consists of the root vectors $\a+m\delta$, $\b+m\delta$ and 
$m\delta$ where $m\geq 1$. 
The corresponding generators can be obtained from $e_{\a,\b}$, 
$f_{\a,\b}$ using the formulas given in Ref.\cite{KT}. 
The algebra $U_q(\widehat{sl}_2)$ is a Hopf algebra with the 
comultiplication $\Delta$ defined by the equations 
\be
\Delta(e_i)=e_i\otimes q^{h_i}+1\otimes e_i, ~~~
\Delta(f_i)=f_i\otimes 1+q^{-h_i}\otimes f_i, ~~~
\Delta(h_i)=h_i\otimes 1+1\otimes h_i.  
\label{co}
\ee
Now we introduce the universal $R$ matrix which is an element in 
$U_q(\widehat{sl}_2)\otimes U_q(\widehat{sl}_2)$. 
This matrix satisfies the following simple relation: 
\be
\Delta^{\prime}(g)R=R\Delta(g), ~~~~~g\in U_q(\widehat{sl}_2), 
\label{delta}
\ee
where $\Delta^{\prime}(g)$ is obtained from $\Delta(g)$ by 
the permutation of the corresponding operators. 
The Hopf algebra is called quasitriangular if the following relations 
are satisfied: 
\be
(\Delta\otimes 1)R=R_{13}R_{23}, ~~~~~(1\otimes\Delta)R=R_{13}R_{12}, 
\label{triang}
\ee
where the notations $R_{12}=R\otimes 1$, $R_{23}=1\otimes R$ are used. 
From the equations (\ref{delta}), (\ref{triang}) it follows that the 
universal $R$- matrix satisfies the Yang-Baxter equation: 
\be 
R_{12}R_{13}R_{23}=R_{23}R_{13}R_{12}.  
\label{YB}
\ee
The explicit expression for the universal $R$- matrix for the case 
of the algebra $U_q(\widehat{sl}_2)$ is given in \cite{KT}. 
For us it is sufficient to establish the following facts about it. 
The universal $R$- matrix can be represented in the form 
$R=\sum_{i}A_{i}\otimes B_{i}\in U_q(b_{+})\otimes U_q(b_{-})$, 
where $b_{\pm}$ are the positive and the negative Borel subalgebras of the 
algebra $\widehat{sl}_2$. More precisely it has the form 
$R=K\sum_{i}A_{i}\otimes B_{i}$, where $K=q^{(h_{\a}\otimes h_{\a})/2}$ 
and the operators $A_i$ and $B_i$ are some polynomials in the generators 
$e_{\a,\b}$ and $f_{\a,\b}$ respectively. 
Once the representations of the quantum group $V_1$, $V_2$ are specified 
the corresponding matrix $R_{12}$ which is the building block in the 
framework of the Quantum Inverse Scattering Method is given by the 
projection of the universal $R$- matrix: $R_{12}=(V_1\otimes V_2)R$.

\vspace{0.2in}

{\bf 3. Representations of quantum group and their correspondence.}

\vspace{0.1in}

Let us introduce the representations of the quantum group which are required for 
the construction of the transfer-matrices and Q- operators according to the 
Quantum Inverse Scattering Method. First consider the representation $V_j(\l)$ 
which is obtained as a combination of Jimbo evaluation homomorphism and the 
standard finite- dimensional highest weight spin $j$ representation of the 
algebra $U_q(sl_2)$. In fact consider the mapping 
\[ 
e_{\a}=f,~~~~f_{\a}=e, ~~~~h_{\a}=-h; 
\]
\be
e_{\b}=\l e,~~~~f_{\b}=\l^{-1}f,~~~~h_{\b}=h; 
\label{ev}
\ee
where $e,f,h$ are the standard generators of $U_q(sl_2)$: 
\[
[h,e]=2e, ~~~~~[h,f]=-2f, ~~~~~[e,f]=\frac{q^{h}-q^{-h}}{q-q^{-1}}
\]  
The highest weight spin $j$ representation is realised in the basis 
$|k\ra$, $k=0,1,\ldots 2j$ in the following way: 
\be
f|k\ra=|k+1\ra,~~~~~ e|k\ra=[k]_q[2j+1-k]_q|k-1\ra, ~~~~~h|k\ra=(2j-2k)|k\ra, 
\label{efh}
\ee
where $[x]_q=(q^{x}-q^{-x})/(q-q^{-1})$. These formulas give the representation 
$V_j(\l)$. We denote the corresponding transfer- matrix by $Z_j(\l)$. 
We also need the representation $V_j^{+}(\l)$, where $j$ is an arbitrary 
complex number, given by the same equations (\ref{ev}),(\ref{efh}) except 
that the basis  $|k\ra$ is now infinite $k=0,1,\ldots\infty$. 
We denote the corresponding transfer matrix by $Z_j^{+}(\l)$. 

To construct the Q- operators we will also need the two representations $V^{\pm}$ 
of the positive Borel subalgebra $U_q(b_{+})$ i.e. the subalgebra given by the 
positive generators $e_{\a,\b},h_{\a,\b}$ and their descendants.  
These representations cannot be extended to the representations of the full 
quantum group. The representations $V^{\pm}$ are build up from the so called 
q-oscillator algebras \cite{BLZ}. We define two different operator algebras. 
The first one consists of the operators $a^{+},a,H_a$ obeying the following 
commutational relations: 
\[
[H_a; a^{+}]=2a^{+},~~~~[H_a; a]=-2a, ~~~~~
qa^{+}a-\frac{1}{q}aa^{+}=\frac{\l}{(q-q^{-1})}. 
\]
The second algebra consists of the operators $b^{+},b,H_b$ obeying the 
following commutational relations: 
\[
[H_b; b^{+}]=2b^{+},~~~~[H_b; b]=-2b, ~~~~~
qbb^{+}-\frac{1}{q}b^{+}b=\frac{\l}{(q-q^{-1})}. 
\]
One can choose the the basis $|k\ra$, $k=0,1,\ldots\infty$, such that the 
action of the operators is: 
\be
a^{+}|k\ra=|k+1\ra, ~~~~~
a|k\ra=\frac{\l}{(q-q^{-1})^2}\left(1-q^{2k}\right)|k-1\ra, ~~~~
H_a|k\ra=2k|k\ra
\label{a}
\ee
for the first case and 
\be
b^{+}|k\ra=|k+1\ra, 
~~~~~b|k\ra=\frac{\l}{(q-q^{-1})^2}\left(1-q^{-2k}\right)|k-1\ra, 
~~~~H_b|k\ra=2k|k\ra 
\label{b}
\ee
for the second case. Then the q-oscillator representations $V^{\pm}$ of the 
subalgebra $U_q(b_{+})$ are defined by the following relations: 
\be
e_{\a}=a^{+},~~~~e_{\b}=a,~~~~h_{\a}=H_a,~~~~h_{\b}=-H_a 
\label{Va}
\ee
for the representation $V^{+}$ and 
\be
e_{\a}=b^{+},~~~~e_{\b}=b,~~~~h_{\a}=H_b,~~~~h_{\b}=-H_b  
\label{Vb}
\ee
for the representation $V^{-}$. 
One can show that the operators corresponding to the auxiliary spaces 
given by the representations $V^{\pm}$ are the Baxter Q- operators $Q^{\pm}(\l)$. 

Now from the equations (\ref{ev}), (\ref{efh}), (\ref{a})-(\ref{Vb}) one can see 
that the representations $V^{\pm}$ can be obtained as a special limits of the 
representations $V_j(\l)$, $V_j^{+}(\l)$ (for example see \cite{N2}). 
In fact one has the equations: 
\be
\lim_{j\to\infty}\l q^{-2j-1}[k]_q[2j+1-k]_q\rightarrow 
\frac{\l}{(q-q^{-1})^2}\left(1-q^{-2k}\right)~~~~~(V^{-}), 
\label{j+}
\ee
\be 
\lim_{j\to-\infty}\l q^{2j+1}[k]_q[2j+1-k]_q\rightarrow 
\frac{\l}{(q-q^{-1})^2}\left(1-q^{2k}\right)~~~~~(V^{+}), 
\label{j-}
\ee
which show that the representations $V^{\pm}$ of the positive Borel subalgebra 
$U_q(b_{+})$ can be obtained as a limits $j\to\pm\infty$ of the representation 
$V_j^{+}$. Note that this correspondence cannot be extended to the representations 
of the full quantum group $U_q(\widehat{sl}_2)$. 
On the level of generators of quantum group the last equations imply the 
following substitution of generators: 
\[
e_{\a}=f\to a^{+}(\l), ~~~~e_{\b}=\l q^{-2j-1}e(j)\to a(\l), ~~~~(V^{+}), 
\]
and 
\[
e_{\a}=f\to b^{+}(\l),~~~~e_{\b}=\l q^{2j+1}e(j)\to b(\l), ~~~~(V^{-}), 
\]
which will be used in the next section.

\vspace{0.2in}

{\bf 4. Tensor product of the representations and functional relations.}

\vspace{0.2in}

Let us derive the functional relations in the universal form the 
representations of the quantum group. 
The simplest way to derive the Baxter $TQ$- equations and the so called 
fusion relations for the product of two transfer matrices is to consider 
the tensor product of two representations $V^{\pm}$. 
For example consider the tensor product $V^{-}\otimes V^{+}$ in detail. 
One should consider the product of the two operators 
$Q^{-}(\l^{\prime\prime})Q^{+}(\l^{\prime})$. For each of the operators 
one has the following formal expression: 
\be
Q^{\pm}(\l)=\tr_{V^{-}\otimes V^{+}}R= 
\sum_{i}\sum_{k=0}^{\infty}q^{kS}\la k|A_{i}(V^{\pm})|k\ra B_i, 
\label{Q} 
\ee
where $B_i$ - are the operators which act in the quantum space and we 
denote by $S$ the value of the generator $-h_{\a}$ in the quantum space 
($S=S^{z}$ for the XXZ- spin chain).  
To evaluate the product of two $Q$- operators one can use the first of the 
relations (\ref{triang}) to get the following expression: 
\be
Q^{-}(\l^{\prime\prime})Q^{+}(\l^{\prime})=
\sum_{i}\sum_{n,l=0}^{\infty}q^{(n+l)S}\la n,l|\Delta(A_{i})(V^{-}\otimes V^{+})
|n,l\ra B_i, 
\label{nl}
\ee
where $|n,l\ra=|n\ra|l\ra$ is the basis in the space $V^{-}\otimes V^{+}$ 
(the operators $a^{+},a$ act in the first space and the operators 
$b^{+},b$- in the second, see section 3). 
Now one should choose the convenient basis in the space $|n,l\ra$. 
Using the equations (\ref{co}) one finds the following expressions: 
\[
\Delta(e_{\a})=a^{+}+b^{+}q^{H_a}, ~~~~~\Delta(e_{\b})=a+bq^{-H_a}. 
\] 
The convenient basis is given by the equation: 
\[
|\phi_{k,m}\ra=(\Delta(e_{\a}))^{k}|m\ra|0\ra.
\]
There exists the operator $U$ such that $|\phi_{k,m}\ra=U|m,k\ra$. 
Inserting $1=UU^{-1}$ into the right-hand side of the equation (\ref{nl}) 
one can represent this equation in the following form:
\be
Q^{-}(\l^{\prime\prime})Q^{+}(\l^{\prime})=
\sum_{i}\sum_{k,m=0}^{\infty}q^{(k+m)S}\la m,k|U^{-1}\Delta(A_{i})
(V^{-}\otimes V^{+})|\phi_{k,m}\ra B_i.  
\label{km}
\ee
The action of the operators $\Delta(e_{\a,\b})$ on the states $|\phi_{k,m}\ra$ 
has the following form: 
\be
\Delta(e_{\a})|\phi_{k,m}\ra=|\phi_{k+1,m}\ra,  ~~~~~~~
\Delta(e_{\b})|\phi_{k,m}\ra=\l [k]_q[2j+1-k]_q |\phi_{k-1,m}\ra + 
c_{k,m}|\phi_{k,m-1}\ra 
\label{action}
\ee
for the spectral parameters $\l^{\prime}=\l q^{-2j-1}$, 
$\l^{\prime\prime}=\l q^{2j+1}$. The explicit form of the constants $c_{k,m}$ 
is not important. The commutativity of $Z_j(\l)$ and $Q^{\pm}(\l)$ follows 
from the universal Yang-Baxter equation (\ref{YB}), while the commutativity of 
$Q^{\pm}(\l)$ follows from the fact that these operators can be obtained 
as a special limits of the transfer matrices (see Section 5). 
From the equations (\ref{km}), (\ref{action}), using the equality 
$U^{-1}|\phi_{k,m}\ra=|m,k\ra$ and the ortogonality of the basis $|n,l\ra$, 
we come to the transfer matrix $Z_j^{+}(\l)$ which has the following formal 
expression: 
\be
Z_j^{+}(\l)=\sum_{i}\sum_{k=0}^{\infty}q^{(k-j)S}\la k|A_{i}(V_j^{+})|k\ra B_i. 
\label{exzj}
\ee
Comparing this equation with the equation (\ref{km}) we obtain the 
following basic functional relation: 
\be
Z_j^{+}(\l)=\frac{1-q^{S}}{q^{jS}}Q^{-}(\l q^{2j+1})Q^{+}(\l q^{-2j-1}). 
\label{zj+}
\ee
It is convenient to redefine the operators $Q^{\pm}(\l)$ according to the 
equations: 
\[
Q^{-}(\l)=\l^{S/4}Q^{\prime -}(\l),~~~~Q^{+}(\l)=\l^{-S/4}Q^{\prime +}(\l). 
\]
Then the relation (\ref{zj+}) takes the form: 
\be 
Z_j^{+}(\l)=CQ^{\prime -}(\l q^{2j+1})Q^{\prime +}(\l q^{-2j-1}), 
~~~~~C=q^{S/2}(1-q^{S}). 
\label{zj++}
\ee
The next step is to find the relation connecting the transfer- matrices 
$Z_j(\l)$ and $Z_j^{+}(\l)$. Taking into account the equations (\ref{efh}), 
(\ref{exzj}) one can see that the sum over $k$ from $2j+1$ to $\infty$ in 
eq.(\ref{exzj}) is equal to $Z_{-j-1}^{+}(\l)$. Thus we obtain: 
\be 
Z_j^{+}(\l)=Z_j(\l)+Z_{-j-1}^{+}(\l),~~~~Z_j(\l)=Z_j^{+}(\l)-Z_{-j-1}^{+}(\l). 
\label{+} 
\ee
From these equations we obtain the following fundamental functional relation: 
\be
Z_j(\l)=C\left(Q^{\prime -}(\l q^{2j+1})Q^{\prime +}(\l q^{-2j-1})-
Q^{\prime -}(\l q^{-2j-1})Q^{\prime +}(\l q^{2j+1})\right). 
\label{zj}
\ee
From the equation (\ref{zj}) one can derive all the other functional relations 
in the universal form. For example taking into account the $Z_{0}(\l)=1$, 
$Z_{1/2}(\l)=Z(\l)$ one can easily obtain the Baxter equation in the 
universal form: 
\be 
Z(\l)Q^{\prime\pm}(\l)=Q^{\prime\pm}(\l q^2)+Q^{\prime\pm}(\l q^{-2}). 
\label{zq}
\ee
It is easy to obtain these relations considering the tensor product 
of the representations $V_{1/2}\otimes V^{\pm}$. 
In a similar way using (\ref{zj}), (\ref{zq}) one can obtain the following 
fusion relations for the transfer matrices: 
\be
Z(\l)Z_{j}(\l q^{-2j-1})=Z_{j-1/2}(\l q^{-2j-2})+Z_{j+1/2}(\l q^{-2j})
\label{zz+}
\ee
and 
\be
Z(\l)Z_{j}(\l q^{2j+1})=Z_{j-1/2}(\l q^{2j+2})+Z_{j+1/2}(\l q^{2j}). 
\label{zz-}
\ee
All the relations presented above are written down in the universal 
form i.e. for an arbitrary integrable model in the quantum space 
$U_q(b_{-})$. The normalization of the operators corresponds 
to the appropriate projection of the universal $R$- matrix. 
In order to write down these operators in the polynomial form 
for the specific model one should calculate the corresponding expressions 
for the universal $R$- matrix \cite{N1}. 
Let us note that the relations (\ref{zz+}), (\ref{zz-}) in the polynomial 
form were obtained for the specific models 
(for example for the XXZ- spin chain) long time ago without the use of the 
quantum groups.

\vspace{0.2in}

{\bf 5. Limits of quantum transfer matrices as Baxter Q- operators.}

\vspace{0.2in}

In this section we derive the limits $j\rightarrow\pm\infty$ 
of the transfer matrices $Z_j(\l)$ and $Z_j^{+}(\l)$ and show their relations 
to the Baxter $Q$- operators. 
Consider the transfer matrix $Z_j^{+}(\l)$. First, one can take the limits 
$j\rightarrow\pm\infty$ in the equation (\ref{zj+}) and take into account 
that $Q^{\pm}(0)=(1-q^S)^{-1}$ 
(in order to avoid singularities at $S=0$ one should introduce the twist 
angle $\phi$, we will put $\phi=0$ everywhere for simplicity and consider the 
case $\phi=0$ in all the final expressions). 
Second, the same result can be obtained 
from the equations (\ref{j+}), (\ref{j-}). In fact the sum over $k$ 
in eq.(\ref{exzj}) is determined by the low values of $k$ (the sum in the 
equation (\ref{Q}) converges). Thus both methods leads to the same result:  
\be
\lim_{j\to\infty}\left(Z_j^{+}(\l q^{-2j-1})q^{jS}\right)=Q^{-}(\l), 
\label{limit+}
\ee
\be
\lim_{j\to-\infty}\left(Z_j^{+}(\l q^{2j+1})q^{jS}\right)=Q^{+}(\l). 
\label{limit-}
\ee
One can argue that in the case $j\rightarrow\infty$ the number $k$ 
can take the values $k\sim 2j$ and the equation (\ref{j+}) breaks down. 
However we see from the equation (\ref{zj+}) that the contributions from the 
regions below the value $k=2j+1$ and above this value cancel each other. 
   Now let us consider the limits of the usual transfer matrix $Z_j(\l)$. 
It is convenient to represent eq.(\ref{zj}) in the following form: 
\[
Z_j(\l)=\frac{1-q^{S}}{q^{jS}}Q^{-}(\l q^{2j+1})Q^{+}(\l q^{-2j-1})-
\frac{1-q^{S}}{q^{(-j-1)S}}Q^{-}(\l q^{-2j-1})Q^{+}(\l q^{2j+1}). 
\]
From this equation it is easy to obtain the following limits: 
\be
\lim_{j\to\infty}\left(Z_j(\l q^{-2j-1})\right)=Q^{-}(\l)-Q^{+}(\l),~~~~~S=0, 
\label{QQ+}
\ee
\[
\lim_{j\to\infty}\left(Z_j(\l q^{-2j-1})q^{(-j-1)S}\right)=-Q^{+}(\l),~~~~~S>0, 
\]
\[
\lim_{j\to\infty}\left(Z_j(\l q^{-2j-1})q^{jS}\right)=Q^{-}(\l),~~~~~S<0, 
\] 
and 
\be
\lim_{j\to-\infty}\left(Z_j(\l q^{2j+1})\right)=Q^{+}(\l)-Q^{-}(\l),~~~~~S=0, 
\label{QQ-}
\ee
\[
\lim_{j\to-\infty}\left(Z_j(\l q^{2j+1})q^{jS}\right)=Q^{+}(\l),  ~~~~~S>0, 
\]
\[
\lim_{j\to-\infty}\left(Z_j(\l q^{2j+1})q^{(-j-1)S}\right)=-Q^{-}(\l),~~~~~S<0.   
\]
Note that the transfer matrix $Z_j(\l)$ at $j<0$ is obtained as an analytical 
continuation of this function at $j=0,1/2,1,\ldots$. 
Consider the limits $j\to\infty$ in more details. First, consider the case 
$S=0$. Clearly the contribution $Q^{-}(\l)$ comes from the low values of 
$k\sim 1$ in the sum over $k$ in the formal equation of the type (\ref{exzj}): 
\be 
Z_j(\l)=\sum_{i}\sum_{k=0}^{2j}q^{(k-j)S}\la k|A_{i}(V_j)|k\ra B_i. 
\label{explicit} 
\ee
The term $-Q^{+}(\l)$ comes from the values $k\sim 2j$ in the sum over $k$ in 
this equation. This fact can be confirmed by the following arguments. 
Consider the equation (\ref{limit+}). The sum over $k$ is infinite for $Z_j^{+}$. 
The contribution of the region 
$k\geq 2j+1$ in eq.(\ref{exzj}) equals $Q^{+}(\l)$. 
In fact, substituting $k=2j+1+k^{\prime}$, $k^{\prime}\geq 0$   
in eq.(\ref{efh}) and taking the limit $j\to\infty$ we obtain exactly the 
operator $Q^{+}(\l)$. Taking into account the equations (\ref{exzj}), 
(\ref{limit+}), (\ref{explicit}) one can see that the contribution of the 
region $k=2j+1-k^{\prime}$, $k^{\prime}>0$, is equal to $-Q^{+}(\l)$ in 
agreement with the equation (\ref{QQ+}). 
One can come to the same conclusion by noticing  
that the contribution of the region above $k=2j+1$ 
is equal to the limit of $Z_{-j-1}^{+}(\l q^{-2j-1})$ 
which can be easily evaluated. Thus the appearance of the term $-Q^{+}(\l)$ 
in the equation (\ref{QQ+}) is explained. 
      One can also calculate this contribution explicitly. 
In fact, taking $k=2j+1-k^{\prime}$, $k^{\prime}\geq 0$, we see that the 
representation of the quantum group corresponding to this contribution is: 
\be
e_{\a}|k^{\prime}\ra=|k^{\prime}-1\ra, ~~~~
e_{\b}|k^{\prime}\ra=\frac{\l}{(q-q^{-1})^2}\left(1-q^{-2k^{\prime}}\right) 
|k^{\prime}+1\ra,  
\label{repr}
\ee
where $k^{\prime}\geq 0$. From eq.(\ref{repr}) one can see that the 
contribution of the $i$-th term in the decomposition 
$R=K\sum_{i}A_{i}\otimes B_{i}$ has the form 
\[
\sum_{N=0}^{N_i}C_N\sum_{k^{\prime}=1}^{\infty}q^{-(S+2N)k^{\prime}}, 
\]
where $C_N$ are some constants and $N_i$ are the number of the generators 
$e_{\b}$ in $A_i$, while the contribution of the same term for the 
operator $Q^{+}(\l)$ equals 
\[
\sum_{N=0}^{N_i}C_N\sum_{k=0}^{\infty}q^{(S+2N)k}. 
\]
Comparing these two contributions one can easily see that the desired 
contribution is in fact equal to $-Q^{+}(\l)$. 
Next one can see that the limits (\ref{QQ+}) for $S>0$ and $S<0$ agree with the 
general formula (\ref{explicit}). The case $S<0$ is obvious while the limit 
in the case $S>0$ follows from the arguments presented above. 
As for the limits (\ref{QQ-}) they are obviously related to the limits 
(\ref{QQ+}) via the relation $Z_j(\l)=-Z_{-j-1}(\l)$. 
Finally one can also see that the limits (\ref{QQ+}), (\ref{QQ-}) are in 
agreement with the relation (\ref{+}). 
While the equations (\ref{limit+}), (\ref{limit-}) where written down in 
\cite{N3}, the equations (\ref{QQ+}), (\ref{QQ-}) - are the main result 
of the present paper. 
We have found the limits of the transfer matrices $Z_{j}(\l)$, 
$Z_{j}^{+}(\l)$ both by the explicit calculations and from the 
functional relations (\ref{zj++}), (\ref{zj}). 
The explicit microscopic calculations presented above are useful for the 
understanding of the results (\ref{limit+}) - (\ref{QQ-}).

Now we show how to obtain the Baxter equation from the fusion relations 
without the knowledge of the explicit form of the $Q$- operators and 
their relation to the q-oscillator algebras. 
The point is that the fusion relations for the transfer matrices 
were obtained without the use of the quantum groups so that the 
Baxter $Q$- operator could be obtained without the knowledge of 
the q-oscillator representations. 
For example consider the fusion relation (\ref{zz+}). 
This equation can be represented in the following form: 
\[
Z(\l)Z_{j}(\l q^{-2j-1})=Z_{j-1/2}(\l q^{-2(j-1/2)-1}q^{-2})+ 
Z_{j+1/2}(\l q^{-2(j+1/2)-1}q^{2}). 
\]
Considering the case $S<0$, multiplying both sides of this equation 
by the factor $q^{jS}$ and taking the limit $j\rightarrow\infty$ we obtain 
the equation: 
\[
Z(\l)Q^{-}(\l)=Q^{-}(\l q^{-2})q^{S/2}+ Q^{-}(\l q^{2})q^{-S/2}. 
\]
It is easy to prove that in all the other cases we get the same equation. 
In the same way from the fusion relation (\ref{zz-}) we obtain 
\[
Z(\l)Z_{j}(\l q^{2j+1})=Z_{j-1/2}(\l q^{2(j-1/2)+1}q^{2})+ 
Z_{j+1/2}(\l q^{2(j+1/2)+1}q^{-2}),  
\]
which in the limit $j\rightarrow-\infty$ leads to the equation: 
\[
Z(\l)Q^{+}(\l)=Q^{+}(\l q^{-2})q^{-S/2}+ Q^{+}(\l q^{2})q^{S/2}. 
\]
These equations are equivalent to the usual equations for the standard 
Baxter $Q$ -operators $Q^{\prime\pm}(\l)$: 
\[
Z(\l)Q^{\prime\pm}(\l)=Q^{\prime\pm}(\l q^2)+Q^{\prime\pm}(\l q^{-2}).
\]
Thus we obtained the Baxter $Q$- operators without the knowledge 
about their connection with the q-oscillator algebras. 
Surely this is connected with the relations of the representations 
of quantum group given in section 3.

\vspace{0.2in}

{\bf 6. Conclusion.}

\vspace{0.2in} 

In conclusion, 
we obtained the Baxter Q-operators in the quantum group 
$U_q(\widehat{sl}_2)$ invariant integrable models as a limits of 
the quantum transfer matrices corresponding to different 
spins in the auxiliary space. 
We showed that the Baxter equation and the form of the 
$Q$- operator follow from the usual fusion functional relations 
for the quantum transfer matrices. 
The main result of the present paper is the explicit 
microscopic calculations of the limits of various transfer- 
matrices. Although these limits can be obtained from the 
known functional relations it seems interesting and important 
to derive them directly by the explicit calculations. 
All the relations are represented in the universal form and 
valid for an arbitrary quantum integrable model in the 
quantum space, for example, for the XXZ- spin chain. 
In order to find the functional relations for the concrete 
model, one should calculate the universal $R$- matrix 
in a given representation \cite{N2}, or to find the $R$- and $L$- 
operators as a polynomials in the spectral parameter.

\end{document}